\documentclass[10pt,conference]{IEEEtran}
\IEEEoverridecommandlockouts

\usepackage{amsmath,amssymb,amsfonts}
\usepackage{algorithmic}
\usepackage{graphicx}
\usepackage{textcomp}

\usepackage[bottom,hang,flushmargin]{footmisc}
\usepackage{graphicx}
\usepackage{subcaption}
\usepackage{colortbl}
\usepackage{xspace}
\usepackage{multirow}
\usepackage{adjustbox}
\usepackage[table,xcdraw]{xcolor}
\usepackage{tabularray}
\usepackage{array}
\usepackage{float}
\usepackage{pifont}
\usepackage{enumitem}
\usepackage{url}
\usepackage{xstring}
\usepackage{bm}
\usepackage{makecell}
\usepackage{algorithm}
\usepackage[normalem]{ulem}
\usepackage{microtype}
\usepackage{listings}
\usepackage{arydshln}
\usepackage{booktabs}
\definecolor{myred}{rgb}{0.6,0,0}
\definecolor{mygreen}{rgb}{0,0.6,0}
\definecolor{mygray}{rgb}{0.5,0.5,0.5}
\definecolor{mymauve}{rgb}{0.58,0,0.82}
\usepackage[
  colorlinks=true,
  linkcolor=mygreen,
  citecolor=myred,
  urlcolor=blue
]{hyperref}
\usepackage[most]{tcolorbox}
\usepackage[compress]{cite}

\newcommand{\dataset}{\textsc{BinAtlas}\xspace}
\newcommand{\vdf}{\textsc{BinAres}\xspace} 
\newcommand{\tool}{\textsc{DeJina}\xspace}

\newcommand*{\figu}{{Figure}\xspace}

\newcommand*{\tabl}{{Table}\xspace}

\newcommand{\add}[1][]{#1}

\newcounter{insightcounter}
\newcommand{\insightnum}{%
	\stepcounter{insightcounter}%
	\theinsightcounter}
\newcommand{\insight}{
    \textbf{Insight \insightnum:}
}

\newcounter{observationcounter}
\newcommand{\observationnum}{%
	\stepcounter{observationcounter}%
	\theobservationcounter}
\newcommand{\observation}{
    \textbf{Observation \observationnum:}
}
    
\begin{document}

\renewcommand{\footnoterule}{
  \vspace*{0.2cm} 
  \hrule width 0.4\columnwidth
  \vspace*{0.2cm} 
}

\title{A Large Scale Study of AI-based Binary Function Similarity Detection Techniques for Security Researchers and Practitioners}
\author{
\IEEEauthorblockN{
    Jingyi Shi\textsuperscript{1,2} \IEEEauthorrefmark{2}\IEEEauthorrefmark{3}, 
    Yufeng Chen\textsuperscript{1,2} \IEEEauthorrefmark{2}\IEEEauthorrefmark{3}, 
    Yang Xiao\textsuperscript{1,2} \IEEEauthorrefmark{1}\IEEEauthorrefmark{2}\IEEEauthorrefmark{3}, 
    Yuekang Li\textsuperscript{3}, 
    Zhengzi Xu\textsuperscript{4},
    Sihao Qiu\textsuperscript{1,2} \IEEEauthorrefmark{2}\IEEEauthorrefmark{3}, \\
    Chi Zhang\textsuperscript{1,2} \IEEEauthorrefmark{2}\IEEEauthorrefmark{3}, 
    Keyu Qi\textsuperscript{1,2} \IEEEauthorrefmark{2}\IEEEauthorrefmark{3}, 
    Yeting Li\textsuperscript{1,2} \IEEEauthorrefmark{2}\IEEEauthorrefmark{3}, 
    Xingchu Chen\textsuperscript{1,2} \IEEEauthorrefmark{2}\IEEEauthorrefmark{3},
    Yanyan Zou\textsuperscript{1,2} \IEEEauthorrefmark{2}\IEEEauthorrefmark{3}, 
    Yang Liu\textsuperscript{5}, 
    Wei Huo\textsuperscript{1,2} \IEEEauthorrefmark{1}\IEEEauthorrefmark{2}\IEEEauthorrefmark{3}
}

\IEEEauthorblockA{
    \textsuperscript{1}Institute of Information Engineering, Chinese Academy of Sciences, Beijing, China \\
    \textsuperscript{2}School of Cyber Security, University of Chinese Academy of Sciences, Beijing, China \\
    \textsuperscript{3}University of New South Wales, Sydney, Australia \\
    \textsuperscript{4}Imperial College London, Imperial Global Singapore, Singapore \\
    \textsuperscript{5}Nanyang Technological University, Singapore \\
    \{shijingyi, chenyufeng, xiaoyang, qiusihao, zhangchi2024, qikeyu, liyeting, chenxingchu, zouyanyan, huowei\}@iie.ac.cn \\
    yuekang.li@unsw.edu.au, z.xu@imperial.ac.uk, yangliu@ntu.edu.sg
}

    \thanks{\IEEEauthorrefmark{1}Corresponding author.}
    \thanks{\IEEEauthorrefmark{2}Key Laboratory of Network Assessment Technology, Chinese Academy of Sciences, Beijing, China}
    \thanks{\IEEEauthorrefmark{3}Beijing Key Laboratory of Network Security and Protection Technology, Beijing, China}
}


\maketitle

\begin{abstract}
Binary Function Similarity Detection (BFSD) is a foundational technique in software security, underpinning a wide range of applications including vulnerability detection, malware analysis. 
Recent advances in AI-based BFSD tools have led to significant performance improvements.
However, existing evaluations of these tools suffer from three key limitations: a lack of in-depth analysis of performance-influencing factors, an absence of realistic application analysis, and reliance on small-scale or low-quality datasets.

In this paper, we present the first large-scale empirical study of AI-based BFSD tools to address these gaps. We construct two high-quality and diverse datasets: \dataset{}, comprising 12,453 binaries and over 7 million functions for capability evaluation; and \vdf{}, containing 12,291 binaries and 54 real-world 1-day vulnerabilities for evaluating vulnerability detection performance in practical IoT firmware settings.
Using these datasets, we evaluate nine representative BFSD tools, analyze the challenges and limitations of existing BFSD tools, and investigate the consistency among BFSD tools.
We also propose an actionable strategy for combining BFSD tools to enhance overall performance (an improvement of 13.4\%). 
Our study not only advances the practical adoption of BFSD tools but also provides valuable resources and insights to guide future research in scalable and automated binary similarity detection.
\end{abstract}

\begin{IEEEkeywords}
Binary Function Similarity Detection, Artificial Intelligence, In-practice Strategy, Dataset and Evaluation
\end{IEEEkeywords}

\maketitle

\section{Introduction}
Binary Function Similarity Detection (BFSD) aims to quantify the similarity between binary functions and has diverse applications in software security, including vulnerability detection~\cite{StatisticalSimilarity,Reoptimization,VulSeeker,Gemini,ASM2Vec,SAFE,Codee,Asteria,AsteriaPro,AIPoweredStudy,Ahmad2023UnsupervisedBC}, malware identification~\cite{MetamorphicMalware,BitshredMalware,Wang2023CanAD}, software composition analysis~\cite{Li2023LibAMAA,Jiang2024BinaryAIBS,Dong2024LibvDiffLV}, and software plagiarism detection~\cite{ProgramLogicPlagiarism,Luo2014SemanticsbasedOB}. 
BFSD serves as a foundational technology in these domains. 
For instance, in vulnerability detection scenarios, users employ BFSD tools to compare functions within target binary programs against known vulnerable functions, thereby identifying 1-day vulnerabilities. 
Given its extensive usage, studying and improving BFSD techniques is of significant importance.

Recently, AI-based BFSD tools~\cite{CLAP, SAFE, HermesSim, Gemini, GMN, ASM2Vec, PALMTree} have demonstrated superior performance by leveraging various models to analyze different representations of binary functions, outperforming traditional BFSD methods~\cite{Eschweiler2016discovREEC,Reoptimization,StatisticalSimilarity,ExtractingConditional,SemanHash}.

To systematically evaluate the performance of AI-based BFSD tools, several studies have been conducted~\cite{CiscoStudy,AIPoweredStudy}. However, they exhibit three key limitations.
First, they lack fine-grained analyses of factors that influence tool performance. 
In real-world scenarios, key factors such as function inlining and function pool size can significantly affect effectiveness. Although previous studies evaluated BFSD tools under several conditions, yet the extent and nature of these impacts remain unclear. A deeper analysis is essential to reveal tool-specific limitations and common challenges.
Second, these studies often overlook practical usage scenarios. In practice, users may combine multiple tools to enhance performance, but the potential and rationale of such combinations have not been explored.
Finally, prior evaluations rely on small-scale datasets with quality issues—including biased project selection, improper compilation settings, and flawed labeling—which introduce systematic bias and limit the generalizability of their conclusions.

In this paper, we address the aforementioned gaps through a large-scale empirical study. To ensure comprehensiveness, we construct two extensive datasets: \dataset{} and \vdf{}.
\dataset{} is designed to evaluate BFSD tools under diverse and realistic conditions. It captures broad real-world variability, comprising 12,453 binaries and 7,339,256 functions compiled from popular projects spanning six categories across 320 distinct compilation configurations.
We ensure its correctness through metadata verification and labeling based on debug information.
\vdf{} is tailored for vulnerability detection experiments. It includes 12,291 binaries and 3,676,923 functions extracted from 58 IoT firmware images. The dataset features 54 known 1-day vulnerabilities as queries and 1,442 manually identified homologous functions as ground truths.
Leveraging these large-scale datasets, our study aims to uncover the limitations and challenges of existing BFSD tools, and to \textbf{gain insights into practical strategies that remain effective as new tools continue to emerge}.

Specifically, we aim to address the following research questions (RQs):

\setlist{nolistsep}
\begin{itemize}[leftmargin=*, noitemsep]
    \item \textit{\textbf{RQ1:} How do different factors impact BFSD tools?}
    \item \textit{\textbf{RQ2:} Can BFSD tools be combined to improve overall results?}
    \item \textit{\textbf{RQ3:} How do BFSD tools and the combination strategy perform in large-scale real-world vulnerability detection?}
\end{itemize}



By addressing the proposed research questions, we gain valuable insights into the inconsistencies among BFSD tools and derive an actionable strategy to enhance their practical applicability. Specifically, we find that tools based on different representations exhibit distinct failure patterns. Building on this observation, we propose a combination strategy that achieves a 13.4\% improvement over the best-performing individual tool in a large-scale real-world vulnerability detection task.

Beyond these practical insights, our study also highlights promising directions for future research. First, our findings suggest that integrating multiple representations within a single tool can further improve overall performance. Second, our evaluation identifies key challenges faced by current BFSD tools, particularly in handling inconsistencies introduced by function inlining and mitigating performance degradation in large-scale settings. Lastly, reducing the manual effort required to verify ranked candidate functions remains an open problem, pointing to the need for more automated or reliable verification mechanisms.

In summary, our contributions are as follows:
\begin{itemize}[leftmargin=*]
    \item \textbf{Datasets:} We present two high-quality and diverse datasets: \dataset, designed for comprehensive capability evaluation, and \vdf, a large-scale dataset tailored for real-world vulnerability detection. Significant human effort was devoted to compilation, verification, and labeling to ensure the accuracy and reliability of both datasets. We publicly release these datasets to support and advance future research in the BFSD community.
    \item \textbf{Large-scale Experiments:} We conduct comprehensive evaluations of nine BFSD tools and present the first in-depth investigation into their effectiveness in realistic usage scenarios.
    \item \textbf{Practical Strategy:} We propose an actionable strategy to improve the effectiveness of BFSD tools by combining BFSD tools, leading to a 13.4\% improvement in a large-scale real-world vulnerability detection task.
    \item \textbf{Future Directions:} We identify key limitations of current BFSD techniques and outline promising future research directions to address the challenges.
\end{itemize}


To facilitate future research, we open-source our source code and dataset here: \url{https://sites.google.com/view/bfsd-study}.

\section{Background}

\begin{figure*}[tb]
    \centering
    \includegraphics[width=0.9\linewidth]{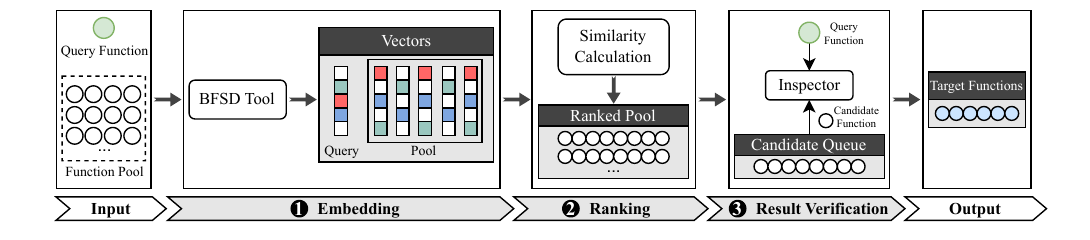}
    \caption{Workflow of a typical BFSD application.}
    \label{fig:workflow}
\end{figure*}

\subsection{Workflow of BFSD Tools}
AI-based function-level BFSD tools transform binary functions into vector representations, casting similarity detection as a vector comparison task.
A typical use case is searching for functions similar to a query within a large-scale pool.
As shown in \figu~\ref{fig:workflow}, this process involves three stages: embedding, ranking, and result verification.

\ding{182} \textbf{Embedding}: The goal of this stage is to convert each input function into a vector representation. BFSD tools extract features from various representations of binary functions and feed them into an AI model to generate embeddings. 

\ding{183} \textbf{Ranking}: 
In this stage, the similarity between the query function’s embedding and each function in the pool is computed to generate a ranked candidate queue. Common similarity metrics include cosine similarity~\cite{Gemini,jTrans,HermesSim,Jiang2024BinColaDC} and Euclidean distance~\cite{Genius}.

\ding{184} \textbf{Result Verification}: 
In the result verification stage, ranked candidates are manually compared to the query function to identify true targets, as the list includes only similarity scores without labels or match guarantees.

\subsection{BFSD Application-oriented Evaluation}
Common applications of BFSD tools include vulnerability detection~\cite{Pewny2014LeveragingSS,Pewny2015CrossarchitectureBS,StatisticalSimilarity,Reoptimization,Genius,VulSeeker,AlphaDiff,Gemini,ASM2Vec,SAFE,Codee,Xu2023PEMRB}, malware identification~\cite{Hu2013MutantXSSM,Cesare2014ControlFM}, software composition analysis~\cite{Li2023LibAMAA,Jiang2024BinaryAIBS,Dong2024LibvDiffLV}, program comprehension~\cite{Hu2016CrossArchitectureBS} and software plagiarism detection~\cite{ProgramLogicPlagiarism,Luo2014SemanticsbasedOB}. These applications share the general workflow illustrated in \figu~\ref{fig:workflow}.
For example, in the widely studied task of vulnerability detection, the common approach is to use a known vulnerable function as the query and search for similar functions in a large pool of binary functions extracted from numerous programs. The goal is to identify vulnerabilities that arise from code reuse or that exist in semantically similar functions.



These real-world scenarios share two main characteristics:
\textbf{Partial knowledge of compilation configurations:} While certain attributes such as architecture and bitness are typically known, other important details—like the specific compiler or optimization level used—are often unavailable~\cite{Chen2022DIComPLD}, which complicates tool performance estimation under varying settings.
\textbf{Large-scale function pools:} In practical applications, function pools often contain millions of functions~\cite{CEBin}. For example, detecting third-party library vulnerabilities in firmware extracted from hundreds of IoT devices involves analyzing a vast number of binary functions.

In practical settings, evaluating BFSD tools and identifying actionable strategies are essential for maximizing their utility.
First, analyzing performance across diverse configurations reveals tool-specific strengths and challenges.
Second, large-scale evaluations set realistic expectations for real-world use.
Finally, exploring external strategies, such as tool combinations, can uncover further performance gains.
Empirical insights into these aspects are key to developing effective and practical solutions.

While previous studies~\cite{CiscoStudy,AIPoweredStudy} have partially evaluated BFSD tools, to the best of our knowledge, none have systematically assessed their performance in large-scale, real-world scenarios or explored effective usage strategies. This paper fills this gap by answering three RQs, offering a comprehensive evaluation and proposing actionable combination strategies to improve real-world effectiveness.
\section{Overview}

\begin{figure*}[htb]
    \centering
    \includegraphics[width=.8\linewidth]{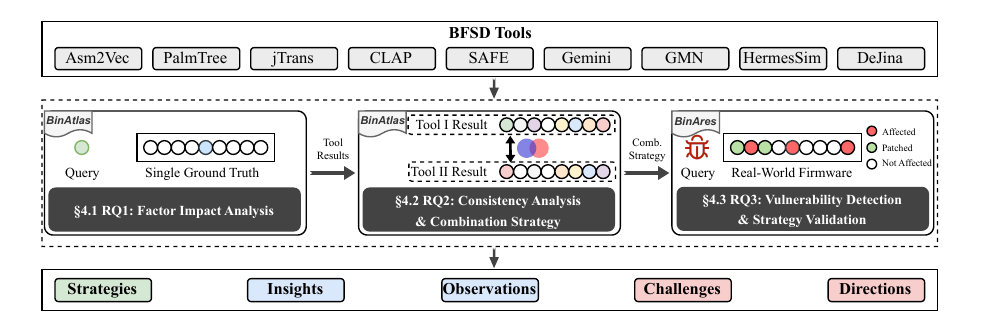}
    \caption{Research questions in this study.}
    \label{fig:rqs}
\end{figure*}

\subsection{Research Questions}
\label{RQs}
In this paper, we investigate three main RQs (as shown in \figu~\ref{fig:rqs}):
\setlist{nolistsep}
\begin{itemize}[leftmargin=*, noitemsep]
    \item \textit{\textbf{RQ1:} How do different factors impact BFSD tools?}
    \item \textit{\textbf{RQ2:} Can BFSD tools be combined to improve overall results?}
    \item \textit{\textbf{RQ3:} How do BFSD tools and the combination strategy perform in large-scale real-world vulnerability detection?}
\end{itemize}

To begin with, in RQ1, we conduct a systematic evaluation of BFSD tools under diverse real-world settings. This allows us to derive results that closely approximate actual performance, and to compare and analyze how different tools perform under various settings, hence identifying the impact of different factors. We also investigate how the key factors influence BFSD tools, both in terms of their nature and the degree of impact.

Building on the findings of RQ1, RQ2 investigates the consistency among different BFSD tools, along with the underlying causes. Motivated by the observed inconsistencies in their failure patterns, we further explore whether combining tools can lead to improved overall performance.

Finally, RQ3 evaluates BFSD tools in a large-scale, real-world vulnerability detection task, with applying the combination strategy from RQ2 to assess its practical effectiveness. The focus is on identifying homologous and vulnerable functions.

Overall, this study aims to systematically uncover the applicability and limitations of BFSD tools.
Based on the insights from the RQs, we propose an effective tool combination strategy, and conclude by outlining practical guidance, key challenges, and future directions in the BFSD domain.

\subsection{BFSD Papers and Tool Selection}
\label{baseline_selection}
BFSD methods can be broadly categorized into dynamic and static approaches. Dynamic methods~\cite{Egele2014BlanketED,Wang2017InmemoryFF, Luo2014SemanticsbasedOB, Xue2019AccurateAS, hu2019semantics, Xu2023PEMRB, zhou2024arcturus} compare the execution semantics of functions through dynamic emulation or execution. These methods tend to be accurate but often lack scalability.
Static methods can be further divided into two subcategories: fuzzy hashing-based methods and AI-based methods.
Fuzzy hashing-based methods~\cite{Reoptimization} map binary functions into fuzzy hashes and compute the similarity between hashes. AI-based approaches extract features from binary functions using raw bytes~\cite{AlphaDiff}, assembly code~\cite{ASM2Vec, SAFE, InnerEye,Trex, PALMTree, jTrans, CLAP, Codee, vexir2vec, BinShot}, decompiled code~\cite{Asteria}, or attribute graphs~\cite{Genius, Gemini, VulSeeker, GraphEmbed, GMN, Sem2Vec, Li2024RCFG2VecCL, he2024strtune, HermesSim, Jiang2024BinColaDC, VulHawk, XBA, OrderMatters, guo2022exploring,Vinyals2015OrderMS} constructed based on dependencies and manually crafted features. These features are then encoded using graph-based or language models to generate vector representations, transforming the function similarity problem into a vector similarity computation, thereby improving efficiency.
Additionally, some techniques enhance the effectiveness of BFSD tools by applying pre-processing to address inconsistencies caused by compilers~\cite{Xu2023ImprovingBC, CodeExtract}, incorporating extra context information~\cite{BinEnhance, AsteriaPro}, re-ranking results~\cite{Wong2024BinAugEB, CEBin,Wang2023EnhancingDB} or adversarial training~\cite{Wang2024ImprovingMB, Jia2024EnhancingLB}.

\begin{table}[htbp]
\caption{Overview of selected BFSD tools. (Rep: Representation, CA: Cross-Architecture, \#Cite: citation count, \#Star: Github stars, \#BL: frequency of use as baselines. A: Assembly, G: Graph, D:Decompiled code.)}

\label{tab:SelectedTools}
\centering
\renewcommand{\arraystretch}{1.2}
\begin{adjustbox}{max width=\linewidth}
\begin{tabular}{lcccccccc}
\hline
\textbf{Tool} & \textbf{Avenue} & \textbf{Year} & \textbf{Rep} & \textbf{Model} & \textbf{CA} & \textbf{\#Cite} & \textbf{\#Star} & \textbf{\#BL} \\ \hline

\rowcolor[rgb]{0.89, 0.89, 0.89}Gemini~\cite{Gemini} & CCS & 2017 & G & Structure2Vec~\cite{Structure2Vec} & \ding{51} & 785 & 135 & 15 \\
GMN~\cite{GMN} & ICML & 2019 & G & GMN~\cite{GMN} & \ding{51} & 736 & 304 & 7\\
\rowcolor[rgb]{0.89, 0.89, 0.89}Asm2Vec~\cite{ASM2Vec} & S\&P & 2019 & A & PV-DM~\cite{PVDM} & \ding{51} & 487 & 624 & 13 \\
PalmTree~\cite{PALMTree} & CCS & 2021 & A & BERT~\cite{Bert} & \ding{55} & 174 & 141 & 7 \\
\rowcolor[rgb]{0.89, 0.89, 0.89}SAFE~\cite{SAFE} & TDSC & 2022 & A & Word2Vec~\cite{word2vec}, SANN~\cite{self-attentive} & \ding{51} & 241 & 175 & 16 \\
jTrans~\cite{jTrans} & ISSTA & 2022 & A & BERT~\cite{Bert} & \ding{55} & 97 & 153 & 6 \\
\rowcolor[rgb]{0.89, 0.89, 0.89}CLAP~\cite{CLAP} & ISSTA & 2024 & A & RoBERTa~\cite{RoBERTa} & \ding{55} & 11 & 54 & 0 \\
HermesSim~\cite{HermesSim} & SEC & 2024 & G & GGNN~\cite{GGNN} & \ding{51} & 16 & 64 & 0 \\
\rowcolor[rgb]{0.89, 0.89, 0.89}\tool & - & - & D & BERT~\cite{Bert} & \ding{51} & - & - & - \\
\hline
\end{tabular}
\end{adjustbox}
\end{table}

In this paper, we limit our scope in AI-based static BFSD approaches at the function level.
We exclude dynamic methods due to their limited scalability in large-scale scenarios. Furthermore, prior work~\cite{CiscoStudy} has demonstrated that recent AI-based BFSD tools outperform earlier fuzzy hashing techniques. Enhancement techniques that can be applied to arbitrary standalone BFSD tools are also excluded from our scope, as they are orthogonal to our focus.

To select BFSD tools for evaluation, we conducted a systematic literature review (SLR). We began by searching leading venues in security, software engineering, AI, and programming languages (e.g., S\&P, CCS, ICSE, FSE, ICML, OOPSLA) for papers published between 2022 and 2024 using keywords such as "binary similarity" and "binary search," yielding 28 papers. After excluding 13 out-of-scope studies through manual screening, we recursively examined the references of the remaining 15 papers. This process resulted in a final set of 34 BFSD papers.
We further selected \textit{representative}, \textit{popular}, and \textit{diverse} BFSD tools for evaluation. First, we excluded ten papers with unavailable tools. Next, we identified six tools published before 2023, based on citation count, GitHub stars, and their use as baselines in prior work. Finally, we included two state-of-the-art (SOTA) tools from 2023 and 2024, based on their GitHub stars. 
The final set of tools is listed in \tabl ~\ref{tab:SelectedTools}.
\add{Our selected tools encompass assembly code-based approaches and graph-based approaches, covering mainstream BFSD methodologies and ensuring diversity.}


The fourth column in \tabl~\ref{tab:SelectedTools} presents the function representations used by the selected BFSD tools, all of which rely on either graph-based or assembly-level  representations. With advancements in source code embedding models, these models can potentially be adapted for BFSD by fine-tuning them for decompiled code similarity detection tasks. Therefore, we fine-tune a source code embedding model, jina-embeddings-v2-base-code\cite{jina-embeddings-v2-base-code}, which is derived from Jina\cite{Jina}, and include it as an additional baseline (\tool).

\subsection{Dataset Construction}
\label{dataset_construction}
To address the three RQs posed in this study, we require two datasets: (1) a representative dataset with diverse real-world configurations to support tool evaluation and consistency analysis in RQ1 and RQ2; and (2) a large-scale, real-world vulnerability detection dataset to assess BFSD tool effectiveness in RQ3 under unknown compilation settings and to validate the strategy proposed in RQ2.

To the best of our knowledge, existing datasets~\cite{jTrans,CiscoStudy,AIPoweredStudy,Zuo2024BinSimDBBD} do not satisfy our requirements.
For the first dataset, existing public datasets suffer from limited diversity, incomplete compilation configurations, and quality issues, making them inadequate for evaluating BFSD tools across varied real-world settings and get reliable results.
For the second dataset, prior vulnerability detection datasets are limited by their small scale, both in terms of the number of vulnerable query functions and the size of the candidate function pool. In addition, their labeling is often incomplete, as annotations typically cover only the top-10 results. These issues hinder reliable and generalizable evaluation. 
Thus, we constructed two datasets for better evaluation: \dataset and \vdf. 


\subsubsection{\dataset}
\dataset\footnote{\dataset~is named after Atlas, a figure from Greek mythology who symbolizes strength and foundation—qualities reflected in this dataset’s comprehensiveness and robustness for BFSD research.} is designed to represent diverse real-world scenarios, guided by three principles: compilation variability, compositional diversity, and correctness. It includes 12,453 binaries compiled from popular projects across six categories using 320 configurations, covering five optimization levels, two compilers (two versions each), four architectures (32/64-bit), and two inlining options (enable function inlining or not). This yields 27.8 million functions, filtered to 7.3 million by excluding short functions that lack meaningful content. \add{Specifically, we exclude functions with fewer than five basic blocks, following the settings of previous papers~\cite{CiscoStudy, Eschweiler2016discovREEC, Genius}.} The dataset is split into training, validation, and test sets without project overlap (as shown in \tabl ~\ref{table:dataset_proj}). \add{The projects are selected based on functionality, popularity, and project size to minimize bias and better reflect real-world scenarios.}
\add{To ensure correctness, we use debug information to label functions with identical names and source positions as positive pairs, avoiding mislabeling from compiler-induced renaming. We also verify binary architecture and optimization levels using the \texttt{file} and \texttt{strings} commands to ensure compilation correctness.} All binaries are stripped after extracting the necessary metadata. 

\begin{table}[htb]
    \caption{Projects used for training, validation, and testing in \dataset. 
Projects marked with * are written in C++, otherwise in C.}
    \label{table:dataset_proj}
    \centering
    \renewcommand{\arraystretch}{1.2} 
    \setlength{\tabcolsep}{4pt} 
    \begin{adjustbox}{max width=\linewidth}
    \begin{tabular}{lcccccc}
        \toprule
        & \textbf{Compression} & \textbf{Network} & \textbf{Text} & \textbf{Database} & \textbf{Image} & \textbf{Other} \\ \midrule
        \rowcolor[rgb]{0.89, 0.89, 0.89} \textbf{Training} & XZ & Nmap*, Openldap, Curl & Xerces-c* & SQLite & ImageMagick & Fmt* \\ 
        \textbf{Validation} & Zlib & Libnet & yaml\_cpp* & - & OpenJPEG & -\\ 
        \rowcolor[rgb]{0.89, 0.89, 0.89} \textbf{Testing} & UnRAR & Openssl, ZeroMQ & JSON* & LevelDB* & Libwebp, Libtiff & PuTTy* \\ 
        \bottomrule
    \end{tabular}
    \end{adjustbox}
\end{table}

\subsubsection{\vdf}
\vdf\footnote{Named after Ares, the Greek god of war, symbolizing strength and challenge in vulnerability detection.} is a large-scale dataset built from 58 real-world firmware images (from 13 vendors including ASUS, Cisco, and Tenda) and 54 known vulnerable functions from nine widely-used libraries \add{(cJSON, Libexpat, LibPNG, Libxml2, Lighttpd1.4, Nginx, OpenSSL, SQLite and Zlib)}. Unlike \dataset, its binaries have unknown compilation settings, reflecting realistic deployment scenarios. 
\vdf comprises 12,291 binaries and 3,676,923 functions across architectures such as 32-bit/64-bit MIPS and 32-bit ARM. The 54 vulnerable functions, compiled under default x86-64 settings, serve as queries. Ground truths were established by manually reviewing functions with the same name and the top-100 results from each tool. \add{Three security experts with at least three years of experience conducted independent annotations, with discrepancies resolved through consensus discussion. This process took over 300 hours and involved inspecting more than 10,000 functions, ultimately identifying 1,442 homologous functions (1–72 per query, median: 23, average: 27).}
 
Details regarding the limitations of prior datasets, as well as comprehensive information on \dataset{} and \vdf{}, are available on our website.

\subsection{Tool Implementation}
\label{baseline_implementation}

All tools, except \tool, have publicly available implementations and require only modifications for adapting to our datasets. For \tool, we fine-tuned jina-embeddings-v2-base-code~\cite{jina-embeddings-v2-base-code}, \add{one of the SOTA source code embedding models available in early 2024,} with a 1024-token limit for efficiency.
CLAP was used in its original zero-shot settings~\cite{CLAP}. 
\tool and other tools were trained with default configurations on the non-inlined subset of \dataset, using a balanced 1:1 ratio of positive and negative function pairs, totaling approximately four million pairs.

\begin{table*}[htb]
\renewcommand{\arraystretch}{1.3} 
\centering
\caption{Performance of BFSD tools across different tasks in \textbf{non-inlined} binaries of \dataset. (R1:Recall@1, R10:Recall@10)}
\label{table:noinline_perf}
\resizebox{.99\textwidth}{!}{ 

\begin{tabular}{cccccccccc}
\Xhline{1pt}
\multirow{2}{*}{\textbf{Tool}} & \textbf{XO} & \textbf{XC} & \textbf{XB} & \textbf{XBCO} & \textbf{XA} & \textbf{XAB} & \textbf{XM} & \textbf{XM$^{RW}$} & \textbf{XM-100k} \\ \cline{2-10}

 & \textbf{R1} / \textbf{R10} / \textbf{MRR} & \textbf{R1} / \textbf{R10} / \textbf{MRR} & \textbf{R1} / \textbf{R10} / \textbf{MRR} & \textbf{R1} / \textbf{R10} / \textbf{MRR} & \textbf{R1} / \textbf{R10} / \textbf{MRR} & \textbf{R1} / \textbf{R10} / \textbf{MRR} & \textbf{R1} / \textbf{R10} / \textbf{MRR} & \textbf{R1} / \textbf{R10} / \textbf{MRR} & \textbf{R1} / \textbf{R10} / \textbf{MRR}  \\ \hline

 \rowcolor[rgb]{0.89, 0.89, 0.89} \textbf{Asm2Vec}
& 0.1 / 0.7 / 0.4 & 0.0 / 0.4 / 0.3 & 0.0 / 0.6 / 0.2 & 0.1 / 0.9 / 0.4 & - / - / - & - / - / - & - / - / - & - / - / - & - / - / - \\

\textbf{PalmTree}
& 42.8 / 50.3 / 45.4 & 41.7 / 52.9 / 45.3 & 2.3 / 6.8 / 3.8 & 24.7 / 29.5 / 26.6 & - / - / - & - / - / - & - / - / - & - / - / - & - / - / - \\

\rowcolor[rgb]{0.89, 0.89, 0.89} \textbf{jTrans}
& 56.9 / 65.7 / 60.1 & 56.8 / 72.2 / 62.3 & 5.8 / 14.3 / 8.9 & 34.6 / 45.9 / 38.3 & - / - / - & - / - / - & - / - / - & - / - / - & - / - / - \\

 \textbf{CLAP}
& 83.8 / 91.0 / 86.2 & 81.6 / 89.0 / 84.3 & 25.4 / 42.1 / 31.2 & 53.8 / 67.2 / 58.3 & - / - / - & - / - / - & - / - / - & - / - / - & - / - / - \\

\rowcolor[rgb]{0.89, 0.89, 0.89} \textbf{SAFE}
& 17.7 / 20.9 / 19.0 & 14.3 / 21.1 / 16.9 & 0.4 / 1.9 / 1.2 & 9.7 / 14.5 / 11.6 & 0.2 / 0.8 / 0.8 & 0.5 / 2.4 / 1.3 & 5.5 / 8.8 / 7.1 & 10.0 / 13.3 / 11.5 & 5.0 / 5.9 / 5.3  \\

 \textbf{Gemini}
& 42.1 / 52.5 / 45.9 & 48.3 / 63.7 / 53.4 & 30.3 / 50.5 / 37.2 & 30.7 / 42.7 / 34.9 & 4.6 / 14.8 / 8.1 & 11.3 / 24.1 / 15.7 & 15.4 / 25.8 / 19.2 & 24.5 / 37.3 / 28.9 & 12.6 / 16.8 / 14.1  \\

\rowcolor[rgb]{0.89, 0.89, 0.89} \textbf{GMN}
& 64.7 / 81.6 / 70.8 & 75.3 / 90.0 / 80.5 & 76.8 / 95.2 / 83.3 & 59.2 / 78.7 / 65.9 & 61.3 / 87.7 / 70.6 & 63.0 / 87.0 / 71.4 & 45.0 / 73.9 / 54.9 & 59.7 / 81.3 / 67.1 & 32.6 / 51.1 / 39.0 \\

\textbf{HermesSim}
& 94.6 / 98.4 / 96.0 &  \textbf{95.9} / \textbf{99.0} / \textbf{97.1} & \textbf{97.7} / \textbf{99.7} / \textbf{98.4} & 94.0 / 97.7 / 95.3 & 93.6 / 98.0 / 95.5 & 93.7 / \textbf{98.5} / \textbf{95.4} & 88.5 / \textbf{95.4} / 91.0 & 89.3 / 94.9 / 91.3 & 83.0 / 92.1 / 86.2 \\

\rowcolor[rgb]{0.89, 0.89, 0.89} \textbf{\tool} 
& \textbf{95.8} / \textbf{98.8} / \textbf{96.8} & 95.8 / 98.6 / 96.8 & 96.1 / 98.9 / 97.1 & \textbf{95.0} / \textbf{98.4} / \textbf{96.2} & \textbf{94.4} / \textbf{98.5} / \textbf{95.9} & \textbf{93.9} / 97.7 / 95.3 & \textbf{90.0} / \textbf{95.4} / \textbf{92.1} & \textbf{91.7} / \textbf{96.3} / \textbf{93.3} & \textbf{84.4} / \textbf{92.6} / \textbf{87.0} \\

\Xhline{1pt}

\end{tabular}
}
\end{table*}

\begin{table*}[h]
\renewcommand{\arraystretch}{1.3} 
\centering
\caption{Performance of BFSD tools across different tasks in \textbf{inlined} binaries of \dataset. (R1:Recall@1, R10:Recall@10)}
\label{table:inline_perf}
\resizebox{.99\textwidth}{!}{ 
\begin{tabular}{cccccccccc}
\Xhline{1pt}
\multirow{2}{*}{\textbf{Tool}} & \textbf{XO} & \textbf{XC} & \textbf{XB} & \textbf{XBCO} & \textbf{XA} & \textbf{XAB} & \textbf{XM} & \textbf{XM$^{RW}$} & \textbf{XM-100k}\\ \cline{2-10}

 & \textbf{R1} / \textbf{R10} / \textbf{MRR} & \textbf{R1} / \textbf{R10} / \textbf{MRR} & \textbf{R1} / \textbf{R10} / \textbf{MRR} & \textbf{R1} / \textbf{R10} / \textbf{MRR} & \textbf{R1} / \textbf{R10} / \textbf{MRR} & \textbf{R1} / \textbf{R10} / \textbf{MRR} & \textbf{R1} / \textbf{R10} / \textbf{MRR} & \textbf{R1} / \textbf{R10} / \textbf{MRR} & \textbf{R1} / \textbf{R10} / \textbf{MRR}  \\ \hline

\rowcolor[rgb]{0.89, 0.89, 0.89} \textbf{Asm2Vec}
& 0.2 / 0.5 / 0.4 & 0.0 / 1.1 / 0.4 & 0.0 / 0.4 / 0.2 & 0.0 / 0.3 / 0.2 & - / - / - & - / - / - & - / - / - & - / - / - & - / - / - \\

\textbf{PalmTree}
& 37.3 / 45.3 / 40.1 & 37.1 / 49.8 / 41.5 & 4.6 / 7.8 / 5.7 & 25.2 / 31.9 / 27.8 & - / - / - & - / - / - & - / - / - & - / - / - & - / - / - \\

\rowcolor[rgb]{0.89, 0.89, 0.89}
\textbf{jTrans}
& 50.3 / 61.0 / 53.7 & 50.9 / 65.7 / 56.1 & 4.2 / 11.4 / 7.0 & 31.1 / 41.3 / 34.7 & - / - / - & - / - / - & - / - / - & - / - / - & - / - / - \\

 \textbf{CLAP}
& 75.0 / 83.3 / 77.9 & 76.5 / 85.2 / 79.5 & 25.4 / 43.4 / 31.2 & 48.5 / 61.3 / 52.8 & - / - / - & - / - / - & - / - / - & - / - / - & - / - / - \\

\rowcolor[rgb]{0.89, 0.89, 0.89} \textbf{SAFE}
& 16.2 / 20.4 / 18.0 & 13.3 / 19.5 / 15.5 & 0.4 / 2.2 / 1.3 & 6.3 / 10.8 / 7.9 & 0.0 / 0.7 / 0.6 & 0.2 / 1.5 / 1.0 & 3.7 / 6.7 / 5.1 & 5.2 / 7.8 / 6.4 & 2.5 / 3.6 / 3.0 \\

\textbf{Gemini}
& 39.2 / 49.8 / 43.0 & 42.0 / 56.2 / 47.2 & 30.1 / 53.1 / 37.7 & 28.7 / 41.2 / 33.0 & 5.1 / 15.9 / 8.9 & 12.3 / 26.3 / 17.0 & 12.3 / 22.3 / 16.3 & 19.2 / 30.0 / 23.2 & 9.7 / 14.7 / 11.3 \\

\rowcolor[rgb]{0.89, 0.89, 0.89} 
\textbf{GMN}
& 58.2 / 73.6 / 63.3 & 65.7 / 83.9 / 71.9 & 79.7 / 94.6 / 85.2 & 52.9 / 72.1 / 59.5 & 66.0 / 87.2 / 73.6 & 66.1 / 88.0 / 73.7 & 42.2 / 63.4 / 49.4 & 49.7 / 69.7 / 56.7 & 28.6 / 44.2 / 34.3 \\

\textbf{HermesSim}
& 88.3 / 93.4 / 90.2  & \textbf{94.6} / \textbf{97.3} / \textbf{95.6} & 96.5 / \textbf{99.3} / 97.6 & 84.3 / 90.9 / 86.7 & 94.7 / \textbf{98.3} / \textbf{96.2} & \textbf{94.8} / \textbf{98.2} / \textbf{96.0} & 81.4 / 89.7 / 84.3 & 85.8 / 93.3 / 88.3 & 75.0 / 83.6 / 77.9 \\

\rowcolor[rgb]{0.89, 0.89, 0.89}  \textbf{\tool}
& \textbf{89.1} / \textbf{94.5} / \textbf{91.0} & 93.6 / 97.1 / 94.8 & \textbf{97.0} / 98.9 / \textbf{97.7} & \textbf{86.9} / \textbf{92.2} / \textbf{88.8} & \textbf{94.8} / 97.3 / 95.8 & 94.4 / 98.1 / 95.8 & \textbf{84.7} / \textbf{91.4} / \textbf{87.2} & \textbf{87.2} / \textbf{93.7} / \textbf{89.4} & \textbf{77.8} / \textbf{85.2} / \textbf{80.5} \\
\Xhline{1pt}

\end{tabular}
}
\end{table*}
\section{Evaluation}\label{evaluation}
In this section, we present each research question along with the corresponding experimental setup, results, and the resulting \textbf{Observations} and \textbf{Insights}. Here, \textbf{Observations} refer to factual and objective findings directly derived from the results, while \textbf{Insights} represent deeper interpretations that can inform the practical application and further development of BFSD tools.

\subsection{Factor Impact Analysis}
\label{RQ1}

\begin{tcolorbox}[colback=white, colframe=black, boxrule=0.75pt, sharp corners, left=5pt,right=5pt,top=5pt,bottom=5pt]
\textbf{RQ1: How do different factors impact BFSD tools?}
\end{tcolorbox}

This RQ explores the performance of diverse BFSD tools across varying compilation settings, aiming to determine their respective applicability and the impact of individual factors.
Specifically, we evaluate each tool across following tasks: XO (cross-optimization levels), XC (cross-compiler and compiler version), XB (cross-bitness), XBCO (a combination of cross-bitness, compiler, compiler version and optimization levels), XA (cross-architecture), XAB (cross-architecture and bitness), XM (cross all compilation configurations), XM$^{RW}$ (a variant of XM that excludes $O0$ and $O1$ optimization levels, as they are rarely used in practice), and XM-100K (an XM variant with an enlarged function pool of 100,000).

\textbf{Experimental Setup.}
For each task, we randomly selected a corresponding positive pair from the test set, designating one function as the query and the other as the ground truth. To construct the function pool, multiple negative samples were randomly chosen for each query, with the ground truth function included. Each tool was then used to compute the similarity scores between the query and all functions in the pool, and to rank the ground truth accordingly. Across all tasks, we randomly selected 1,000 query functions for evaluation.
The pool size for all tasks, except XM-100K, is set to 10,000, following the standard setup in recent studies~\cite{CLAP, HermesSim, CiscoStudy}. For tasks that do not involve cross-architecture comparisons, functions are selected only from the x86 architecture, as Asm2Vec, PalmTree, jTrans and CLAP all support x86.
We adopt Recall@1 (R1), Recall@10 (R10), and mean reciprocal rank (MRR) as evaluation metrics, consistent with prior work~\cite{CLAP, HermesSim, CiscoStudy}.


\textbf{Results.} 
\tabl~\ref{table:noinline_perf} and \tabl~\ref{table:inline_perf} present the performance of each tool on non-inlined binaries and inlined binaries, respectively.
The results indicate that HermesSim and \tool consistently achieved the highest performance, outperforming the next best tool by at least 10 percentage points.
In the XO and XC tasks, CLAP performed slightly below HermesSim and \tool but significantly outperformed all other tools, with GMN ranking next. Conversely, in the XB and XBCO tasks, GMN surpassed Gemini, CLAP, jTrans, PalmTree, SAFE, and Asm2Vec, indicating that both CLAP and GMN exhibit task-specific strengths.
Among assembly code-based approaches, CLAP outperformed all others. In contrast, among graph-based methods, HermesSim achieved the best overall results.

\begin{tcolorbox}[colback=black!6!white,enhanced,frame hidden, boxsep=0pt,left=5pt,right=5pt,top=2pt,bottom=2pt]
\observation
HermesSim and \tool consistently delivered strong performance across all configuration scenarios. CLAP slightly trailed them in the XO and XC tasks, while GMN followed closely in the XB, XA, and XAB tasks.
\end{tcolorbox}

We further examined the performance decline of SAFE, PalmTree, jTrans, and CLAP in the XB and XBCO tasks. These tools primarily focus on addressing the impact of optimization levels on BFSD, with CLAP pre-trained exclusively on binaries differing in optimization levels and compilers. However, cross-bitness configurations introduce substantial changes in register naming conventions and addressing modes, which significantly alter the assembly code structure and degrade the effectiveness of these tools. This highlights a key limitation of assembly-based representations in handling architectural and bitness diversity.

\begin{tcolorbox}[colback=blue!5!white,enhanced,frame hidden, boxsep=0pt,left=5pt,right=5pt,top=2pt,bottom=2pt]
\insight
BFSD tools that use assembly code as representation tend to underperform in the XB and XA tasks, due to limited robustness against syntactic variations introduced by changes in bitness or architecture.
\end{tcolorbox}

When comparing tasks involving cross-single compilation settings across non-inlined dataset and inlined dataset, the XO and XC tasks show noticeable performance degradation, whereas the XB and XA tasks are less affected. This aligns with the understanding that different optimization levels apply varying inlining strategies, which also differ across compilers and evolve with compiler versions. 
To assess the impact of function inlining on BFSD tool performance, we conduct an in-depth analysis of the XO task, where inlining-induced degradation is most evident. In the following analysis, a failure case is defined as one in which the ground truth is not ranked in the top-10 predictions.

We first compare the proportion of inlined functions among all queries versus failure cases for each tool (see "Prop. Inline" in \tabl~\ref{tab:inlining_analysis}). Notably, high-performing tools such as CLAP, GMN, HermesSim, and \tool exhibit a significantly higher proportion of inlined functions in their failure cases, suggesting inlining as a key factor. Lower-performing tools do not exhibit such features, as they are more sensitive to diverse compilation variations, with inlining contributing secondarily to their errors.

To further elucidate the impact of inlining, we distinguish between identical inlining, where both functions in a positive pair inline the same callees, and differential inlining, where the inlined callees differ between the pair. As shown in the "Prop. Inline-Id." and "Prop. Inline-Diff." rows of \tabl~\ref{tab:inlining_analysis}, failure cases for HermesSim and \tool exhibit 0\% identical inlining but over 80\% differential inlining. This indicates that performance degradation in these tools is primarily driven by differential inlining, which introduces semantic inconsistencies in positive function pairs.
To quantify this semantic divergence caused by inlining, we compute the Diff. Ratio, defined as the proportion of instructions originating from non-overlapping inlined callees between a pair of functions. For instance, given a positive function pair $F$ and $F'$, if $F$ inlines callees $G$ and $H$ while $F'$ inlines $G'$ and $I'$, the Diff. Ratio is computed as $\frac{|H| + |I'|}{\text{average}(|F|, |F'|)}$, where $F$ denotes the size of $F$. Here, $G$ and $G'$ are homologous and thus excluded from the calculation.
The last row in \tabl~\ref{tab:inlining_analysis} reports the average Diff. Ratio for failure cases involving differential inlining for each tool.
Across all queries, the average Diff. Ratio is 24.0\%. However, in failure cases, this value rises to 72.9\% for HermesSim and 78.4\% for \tool, indicating that these tools are robust to minor inlining-induced variations but fail when semantic shifts become substantial. In contrast, lower-performing tools exhibit smaller Diff. Ratios in their failure cases, suggesting they are more vulnerable to even slight semantic inconsistencies introduced by inlining.

\begin{table*}[h]
\centering
\caption{Function inlining impact analysis across the failure cases of different tools and queries.}
\label{tab:inlining_analysis}
\begin{tabular}{l|ccccccccc|c}
\hline
\textbf{Metric} & \textbf{Asm2Vec} & \textbf{PalmTree} & \textbf{jTrans} & \textbf{CLAP} & \textbf{SAFE} & \textbf{Gemini} & \textbf{GMN} & \textbf{HermesSim} & \textbf{\tool} & \textbf{Queries} \\
\hline
\rowcolor[rgb]{0.89, 0.89, 0.89} \textbf{Prop. Inline} & 51.6\% & 48.5\% & 49.5\% & 65.3\% & 51.6\% & 52.0\% & 62.9\% & 81.8\% & 83.6\% & 51.5\% \\
\textbf{Prop. Inline-Id.} & 18.6\% & 5.4\% & 3.3\% & 4.8\% & 13.4\% & 5.6\% & 3.0\% & 0.0\% & 0.0\% & 18.7\% \\
\rowcolor[rgb]{0.89, 0.89, 0.89} \textbf{Prop. Inline-Diff.} & 33.0\% & 43.1\% & 46.2\% & 60.5\% & 38.2\% & 46.4\% & 59.8\% & 81.8\% & 83.6\% & 32.8\% \\
\textbf{Diff. Ratio} & 24.1\% & 35.4\% & 38.8\% & 45.9\% & 28.8\% & 35.9\% & 46.9\% & 72.9\% & 78.4\% & 24.0\% \\
\hline
\end{tabular}
\end{table*}

\begin{tcolorbox}[colback=blue!5!white,enhanced,frame hidden, boxsep=0pt,left=5pt,right=5pt,top=2pt,bottom=2pt]
\insight
Function inlining remains a major challenge for BFSD tools. Performance degradation primarily occurs when positive pairs undergo asymmetric inlining. Higher-performing tools, such as HermesSim and \tool, are more robust to function inlining, failing when inlining introduces substantial semantic differences (averaging over 70\%).
\end{tcolorbox}

\begin{figure}[htb]
    \centering
    \includegraphics[width=.75\linewidth]{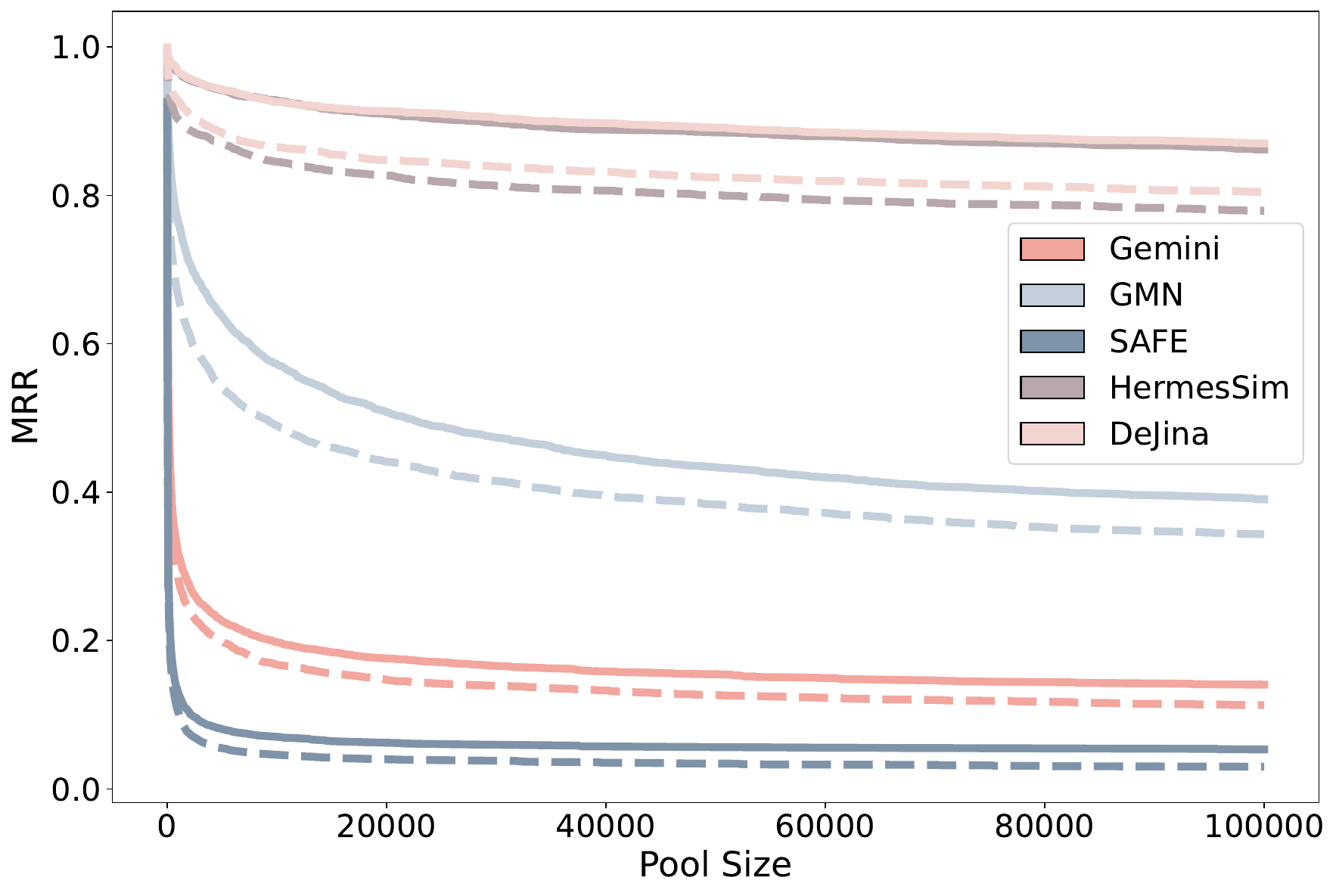}
    \caption{MRR  by the size of function pool. (Solid lines: non-inlined dataset; dashed lines: inlined dataset.)}
    \label{fig:recall_pool}
\end{figure}

A comparison of the XM and XM-100K columns in \tabl~\ref{table:noinline_perf} and \tabl~\ref{table:inline_perf} reveals that as the function pool size increases from 10,000 to 100,000, the overall MRR of all tools declines, with absolute drops ranging from 2.1\% to 15.1\%. However, the extent of degradation varies significantly across tools. HermesSim and \tool demonstrate stronger robustness, with MRR reductions of less than 7\%, whereas Gemini, GMN, and SAFE suffer more pronounced performance drops, each exceeding 25\%, relatively.
\figu~\ref{fig:recall_pool} shows the MRR performance of Gemini, GMN, SAFE, HermesSim, and \tool across increasing function pool sizes. Other tools are excluded due to incompatibility with cross-architecture settings. As observed, MRR drops sharply when the pool size increases from 0 to 20,000, revealing a key limitation of current BFSD evaluations—namely, the lack of assessment under large-scale function pools. Most existing studies cap the pool size at 10,000, which may not reflect real-world conditions.
Although the decline slows as the pool grows, performance continues to degrade. In practical scenarios, where function pools can reach millions, this degradation becomes much more substantial compared to smaller-scale experimental settings.


\begin{tcolorbox}[colback=black!6!white,enhanced,frame hidden, boxsep=0pt,left=5pt,right=5pt,top=2pt,bottom=2pt]
\observation
The size of the function pool affects the performance of BFSD tools. As the pool size increases, their MRR tends to decline rapidly at first and then levels off. 
\end{tcolorbox}

\subsection{Consistency Analysis and Combination Strategy}
\label{RQ2}

\begin{tcolorbox}[colback=white, colframe=black, boxrule=0.75pt, sharp corners, left=5pt,right=5pt,top=5pt,bottom=5pt]
\textbf{RQ2: Can BFSD tools be combined to improve overall results?}
\end{tcolorbox}

\begin{figure}[htb]
    \centering
    \includegraphics[width=0.9\linewidth]{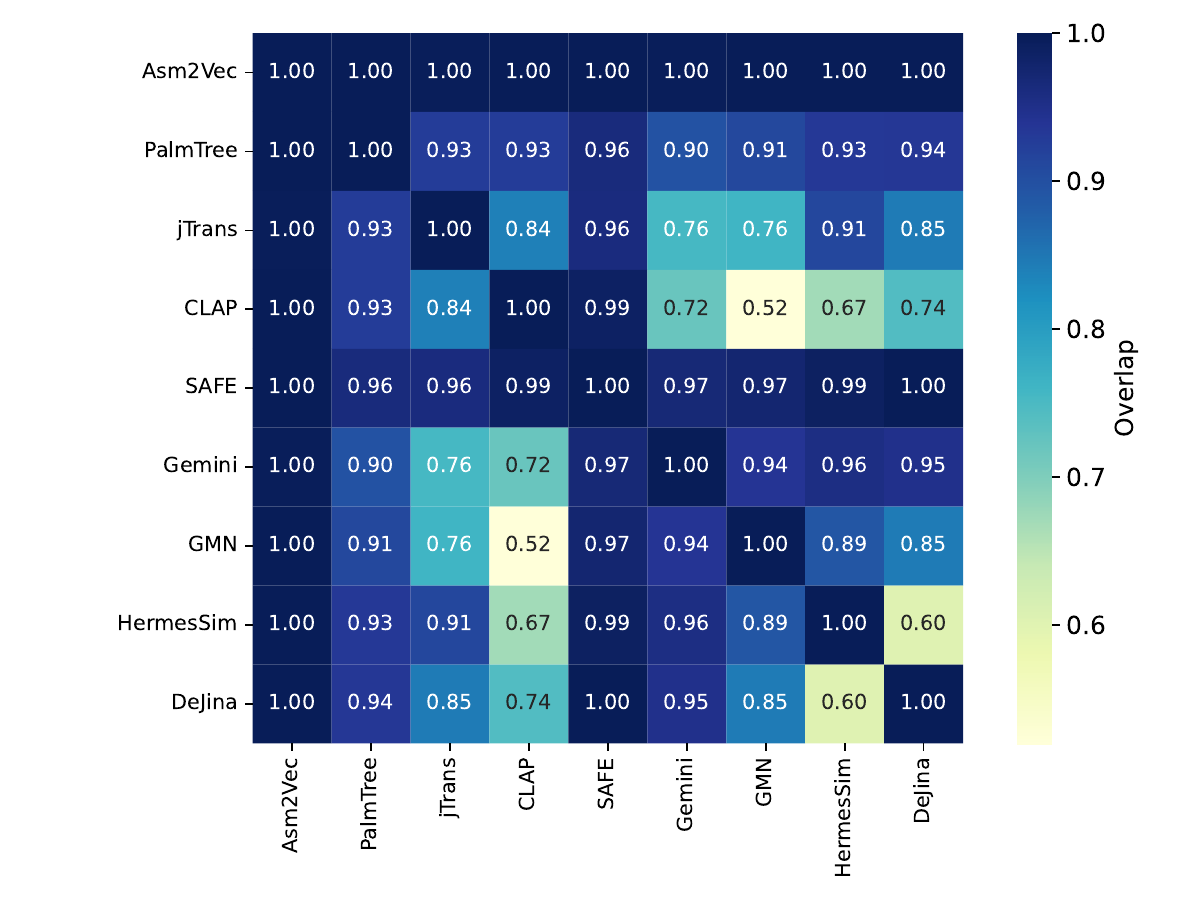}
    \caption{The overlap coefficient between tool pairs.}
    \label{fig:consistency}
\end{figure}

The results of RQ1 reveal that the performance of different tools varies across tasks.
These disparities raise the question of whether the tools can complement one another.
To explore this possibility, this RQ first analyzes the consistency among these tools based on the results from RQ1.
Subsequently, we investigate various combination strategies and evaluate their effectiveness in enhancing overall performance.

\textbf{Experimental Setup.}
In this RQ, a query is considered a failure case if its ground truth is ranked outside the top 10. We analyze the consistency of failure cases between two tools using overlap coefficient~\cite{mcgill1979evaluation}:
$$ overlap(A, B) = \frac{ \lvert A \cap B \rvert}{min(\lvert A \rvert,\lvert B \rvert)}$$
where $A$ and $B$ denote the sets of failure cases for each tool. 
The overlap coefficient quantifies the extent to which the failure cases of the better-performing tool overlap with those of another, with higher values indicating similar failure patterns and reduced potential for complementary gains.

\begin{table*}[htb]
\caption{Reasons for failure cases in HermesSim and \tool.}
\label{tab:FPFN_reason}
\centering
\renewcommand{\arraystretch}{1.3}
\resizebox{0.95\linewidth}{!}{
\begin{tabular}{cccccc}
\hline
\textbf{Tool} & \textbf{Function Inlining} & \textbf{Flow Changes} & \textbf{Indistinctive Functions} & \textbf{Distinct Instruction Sets} & \textbf{Decompiled Code Issues} \\ \hline
\rowcolor[rgb]{0.89, 0.89, 0.89} \textbf{Common} & 42 (89.4\%) & 5 (10.6\%) & 0 (0\%) & 0 (0\%) & 0 (0\%) \\ 
\textbf{HermesSim} & 20 (45.5\%) & 17 (38.6\%) & 7 (15.9\%) & 0 (0\%) & 0 (0\%) \\ 
\rowcolor[rgb]{0.89, 0.89, 0.89} \textbf{\tool} & 15 (48.4\%) & 6 (19.4\%) & 2 (6.5\%) & 4 (12.9\%) & 4 (12.9\%) \\ \hline
\end{tabular}
}
\end{table*}

\textbf{Results.} We analyze the consistency of failure cases across tools on the XBCO task using the inlined dataset. Specifically, we compute the overlap coefficient for all tool pairs, as presented in \figu~\ref{fig:consistency}.
All pairs exhibit coefficients above 0.5, indicating that at least half of one tool's failure cases are shared with the other. But some of them exibit relatively lower value than other tool pairs, e.g. GMN and CLAP (0.52), HermesSim and \tool (0.6). 

\begin{tcolorbox}[colback=black!6!white,enhanced,frame hidden, boxsep=0pt,left=5pt,right=5pt,top=2pt,bottom=2pt]
\observation
Some tool pairs exhibit relatively low overlap, indicating potential for performance improvement through complementary use.
\end{tcolorbox}

To investigate the reason behind their relatively lower inconsistency, we manually analyzed all common (47) and unique ones in HermesSim (44) and \tool (31), as they are the best-performing tools. The reasons are concluded in \tabl ~\ref{tab:FPFN_reason}.

In the table, \ding{182} Function Inlining refers to cases where aggressive inlining by the compiler leads to substantial difference between function pairs, thereby complicating similarity detection.
\ding{183} Flow Changes denote instances where aggressive compiler optimizations significantly modify the control flow structure, resulting in pronounced structural variations in the decompiled code. 
\ding{184} Indistinctive Functions capture scenarios where the target functions lack distinctive characteristics, or where retrieved candidates exhibit high similarity to the query function, making precise differentiation challenging. 
\ding{185} Distinct Instruction Sets describe cases where compilers, driven by performance or size constraints, selectively utilize SIMD instructions instead of conventional ones, leading to significant discrepancies in the decompiled representation. Finally, 
\ding{186} Decompiled Code Issues encompass failures caused by decompiler errors that fail to generate valid decompiled code.

Though HermesSim and \tool share similar underlying causes for their respective unique failures, such as function inlining and flow changes, our analysis reveals that the primary reasons one tool may fail while the other succeeds are rooted in their distinct design philosophies. HermesSim constructs graphs based on intra-function dependency relationships and relies on a graph-based model for feature extraction. Its sensitivity to flow variations allows it to effectively distinguish between functions with similar functionality but different structural patterns. However, this same sensitivity makes it more prone to failure when the flow undergoes substantial transformations, such as those introduced by aggressive compiler optimizations or inlining.
In contrast, \tool leverages language models to generate embeddings from decompiled code, enabling it to extract rich semantic features from symbol names and string literals. This enables it to distinguish functions with similar structural patterns but different symbol names or string literals. However, this reliance on semantic content renders it significantly more vulnerable to the quality and consistency of the decompiler's output. For example, decompilation errors or inconsistent recovery of symbolic information and strings across function pairs can lead to failures in \tool's predictions. Additionally, the insensitivity to dependency relationships makes it fail to distinguish similar functions. A case study for further explaining the inconsistency of HermesSim and \tool is provided on our website.

These observations suggest that BFSD tools built on different representations (graph, assembly code, and decompiled code, etc.) may exhibit complementary strengths and vulnerabilities.

\begin{tcolorbox}[colback=blue!5!white,enhanced,frame hidden, boxsep=0pt,left=5pt,right=5pt,top=2pt,bottom=2pt]
\insight
BFSD tools based on different representations exhibit varying robustness to function pair discrepancies, resulting in inconsistent failure patterns.
This variability highlights the potential for performance improvements through integrating complementary approaches.
\end{tcolorbox}

\textbf{Experimental Setup.}
Motivated by the above observations and insights, we conduct a preliminary investigation into combining BFSD tools to enhance overall performance. The central hypothesis is that tool pairs built on different representations and with comparable performance levels are more likely to be complementary. Tools based on different representations exhibit distinct failure patterns, thereby increasing the likelihood of complementary strengths. At the same time, combining tools with similar performance levels helps mitigate the risk of performance degradation caused by noise from a substantially weaker tool. To evaluate this hypothesis, we assess all pairwise combinations on the XBCO task and examine the resulting performance improvements. 

\textbf{Combination strategy.} To integrate multiple BFSD tools, we propose a voting strategy, which ranks candidate functions based on consensus across tools. Functions retrieved by all tools are prioritized and ranked by their average position, followed recursively by those retrieved by fewer tools. 
This strategy is practical, as it requires no parameter tuning and is compatible with any BFSD tool. For example, if Tool I returns A, B, C, D and Tool II returns D, A, E, C, the overlapping functions A, C, and D are ranked first by average rank: A (1.5), D (2.5), C (3). The remaining functions B and E are then ranked as B (2), E (3), resulting in a final order of A, D, C, B, E.


\begin{table}[htbp]
\renewcommand{\arraystretch}{1.3}
\centering
\caption{The results of tool combinations ranked by MRR improvement. ($\delta$MRR: The MRR gap between the tools in a pair.)}
\label{tab:combo_comparison_all}
\resizebox{\linewidth}{!}{
\begin{tabular}{rccc}
\Xhline{1pt}
\textbf{Combination} & \textbf{$\delta$MRR} & \textbf{Combination Results (R1 / R10 / MRR)} \\
\hline
\rowcolor[rgb]{0.89, 0.89, 0.89} GMN+CLAP & 6.7 & 66.2 (\textcolor{myred}{$\uparrow$}13.3) / 83.5 (\textcolor{myred}{$\uparrow$}11.4) / 72.3 (\textcolor{myred}{$\uparrow$}12.8) \\
jTrans+Gemini & 1.7 & 37.0 (\textcolor{myred}{$\uparrow$}5.9) / 52.8 (\textcolor{myred}{$\uparrow$}11.5) / 43.2 (\textcolor{myred}{$\uparrow$}8.5) \\
\rowcolor[rgb]{0.89, 0.89, 0.89} CLAP+Gemini & 19.8 & 50.8 (\textcolor{myred}{$\uparrow$}2.3) / 69.7 (\textcolor{myred}{$\uparrow$}8.4) / 58.0 (\textcolor{myred}{$\uparrow$}5.2) \\
PalmTree+Gemini & 5.2 & 30.6 (\textcolor{myred}{$\uparrow$}1.9) / 44.1 (\textcolor{myred}{$\uparrow$}2.9) / 35.9 (\textcolor{myred}{$\uparrow$}2.9) \\
\rowcolor[rgb]{0.89, 0.89, 0.89} HermesSim+\tool & 2.1 & 88.7 (\textcolor{myred}{$\uparrow$}1.8) / 94.7 (\textcolor{myred}{$\uparrow$}2.5) / 90.8 (\textcolor{myred}{$\uparrow$}2.0) \\
CLAP+jTrans & 18.1 & 47.5 (\textcolor{mygreen}{$\downarrow$}1.0) / 64.8 (\textcolor{myred}{$\uparrow$}3.5) / 53.9 (\textcolor{myred}{$\uparrow$}1.1) \\

\hline
\rowcolor[rgb]{0.89, 0.89, 0.89} PalmTree+jTrans+SAFE & 26.8 & 26.9 (\textcolor{mygreen}{$\downarrow$}4.2) / 42.0 (\textcolor{mygreen}{$\downarrow$}0.7) / 32.8 (\textcolor{mygreen}{$\downarrow$}2.1) \\              
CLAP+jTrans+Asm2Vec & 52.6 & 44.6 (\textcolor{mygreen}{$\downarrow$}3.9) / 63.3 (\textcolor{mygreen}{$\downarrow$}1.5) / 51.6 (\textcolor{mygreen}{$\downarrow$}2.3) \\               
\rowcolor[rgb]{0.89, 0.89, 0.89} DeJina+HermesSim+GMN & 29.3 & 85.1 (\textcolor{mygreen}{$\downarrow$}3.6) / 94.2 (\textcolor{mygreen}{$\downarrow$}0.5) / 88.4 (\textcolor{mygreen}{$\downarrow$}2.4) \\  

\Xhline{1pt}

\end{tabular}
}
\end{table}

\textbf{Results.}
We list the combinations of two tools that with more than 1\% improvement in MRR in the first part of \tabl~\ref{tab:combo_comparison_all}, the results of all combinations can be found in our website.
Notably, combining GMN and CLAP yields the most consistent improvements across metrics under both strategies.It improves R1, R10, and MRR by 13.3\%, 11.4\%, and 12.8\%, respectively.
The combination of \tool and HermesSim achieves the highest overall performance: 88.7\%, 94.7\%, and 90.8\%.

\begin{tcolorbox}[colback=black!6!white,enhanced,frame hidden, boxsep=0pt,left=5pt,right=5pt,top=2pt,bottom=2pt]
\observation
The proposed combination strategy can yield performance improvements of up to 12.8\%.
\end{tcolorbox}

Moreover, all tool pairs yielding MRR improvements are based on different representations, except CLAP and jTrans, which also show the smallest gain. Effective combinations typically exhibit small performance gaps between paired tools. Notably, all tool pairs with distinct representations and low performance gaps demonstrate improved performance after combination. Conversely, tool pairs with large performance gaps consistently suffer degradation. \add{For example, combining HermesSim and CLAP results in a 6.0\% MRR drop due to their large performance gap (33.9\% $\delta$MRR). These findings empirically validate our hypothesis and establish clear criteria on performance gap for effective tool complementarity.}

\begin{tcolorbox}[colback=blue!5!white,enhanced,frame hidden, boxsep=0pt,left=5pt,right=5pt,top=2pt,bottom=2pt]
\add{\insight}
Effective tool combination requires two criteria: \ding{182} tools must be based on different binary function representations, leading to complementary failure patterns, and \ding{183} tools must exhibit a small performance gap, as large disparities introduce noise from the weaker tool. 
\end{tcolorbox}

Our results show that when the $\delta$MRR exceeds 20\%, combinations fail to improve performance. All tool pairs satisfying both criteria achieve MRR improvements (up to 12.8\%), demonstrating the importance of balancing complementarity with comparable capability.

To further explore potential gains from incorporating additional tools, we iteratively introduce a third tool into the pairwise combinations that previously demonstrated performance improvement. The best-performing three-tool combinations are presented in the second part of \tabl~\ref{tab:combo_comparison_all}. However, all such combinations result in performance degradation. This outcome likely stems from a violation of the criteria, as each three-tool combination includes either two tools based on the same representation or at least with a $\delta$MRR exceeding 20\%.


\subsection{RQ3: Vulnerability Detection \& Strategy Validation}
\label{RQ3}

\begin{tcolorbox}[colback=white, colframe=black, boxrule=0.75pt, sharp corners, left=5pt,right=5pt,top=5pt,bottom=5pt]
\textbf{RQ3: How do BFSD tools and the combination strategy perform in large-scale real-world vulnerability detection?}
\end{tcolorbox}

A typical usage of BFSD is to use a known vulnerable function as the query and search for similar functions within a large binary function pool. This enables the discovery of vulnerabilities that arise from code reuse or that exist in semantically similar functions.
Since the function pool often comprises thousands of binaries from diverse projects, it is common that there are either multiple target functions or none at all, making the number of target functions inherently unknown.

In this RQ, we evaluate the practical effectiveness of the combination strategy proposed in RQ2 on \vdf, a large-scale, real-world dataset. We also highlight key challenges and insights in applying BFSD tools to real-world vulnerability detection. Given \vdf’s multi-architecture nature, only cross-architecture tools are included. Due to the dataset's scale and high evaluation cost, we focus on the top three tools—\tool, HermesSim, and GMN—as well as the best-performing combination from RQ2 (HermesSim + \tool).

\begin{table}[htp]
\centering
\caption{Results of homologous function detection on \vdf. The content in each cell is Precision / Recall / F1 score. (T: Threshold of similarity.)}
\label{tab:firmware}
\renewcommand{\arraystretch}{2}
\resizebox{\linewidth}{!}{
\begin{tabular}{c|c|c|c|c}
\Xhline{1pt}
\textbf{Baseline} & \textbf{Top-10} & \textbf{Top-25} & \textbf{Top-50} &  \textbf{Best Threshold} \\ \hline
\rowcolor[rgb]{0.89,0.89,0.89} GMN & \Gape[0pt][2pt]{26.9 / 9.2 / 13.1} & \Gape[0pt][2pt]{19.6 / 16.0 / 16.6} & \Gape[0pt][2pt]{14.3 / 22.5 / 16.4} & \Gape[0pt][2pt]{\makecell[c]{19.4 / 11.9 / 12.3 \\(T=0.98)}}  \\ 

HermesSim & \Gape[0pt][2pt]{71.3 / 45.4 / 42.7} & \Gape[0pt][2pt]{55.6 / 67.8 / 51.6} & \Gape[0pt][2pt]{36.7 / 78.5 / 43.8} & \Gape[0pt][2pt]{\makecell[c]{45.9 / 43.4 / 40.2 \\ (T=0.49)}}  \\ 

\rowcolor[rgb]{0.89,0.89,0.89} \tool & \Gape[0pt][2pt]{70.2 / 43.3 / 40.5} & \Gape[0pt][2pt]{55.9 / 65.7 / 50.1} & \Gape[0pt][2pt]{38.9 / 80.7 / 45.7} & \Gape[0pt][2pt]{\makecell[c]{\makecell[c]{57.1 / 44.0 / 44.6 \\ (T=0.79)}}}  \\ 

\hline
\makecell[c]{\tool + \\ HermesSim} & \Gape[0pt][2pt]{77.4 / 48.7 / 46.2} & \Gape[0pt][2pt]{63.6 / 75.9 / 58.5} & \Gape[0pt][2pt]{45.0 / 91.9 / 53.4} & -  \\

\Xhline{1pt}

\end{tabular}
}
\end{table}

\textbf{Experimental Setup.}
We use 54 vulnerable functions as queries against a pool of 3,676,923 functions in \vdf, and evaluate BFSD tools across three tasks:
\ding{182} Homologous function detection:
We treat functions derived from the same source code as ground truth. The ground truth was established manually as described in \S ~\ref{dataset_construction}.
\ding{183} Vulnerable Homologous Function Detection:
To evaluate the tools’ effectiveness in identifying vulnerable instances among homologous functions, we manually inspected 500 randomly sampled homologous candidates retrieved by the top-performing tool. This step accounts for scenarios where some homologous functions may have been patched or originate from versions preceding the vulnerability.
\ding{184} Vulnerable Non-Homologous Function Detection:
To assess the ability of tools to identify vulnerabilities beyond strict source-level similarity, we also examined 500 top-ranked non-homologous functions. Each function was manually inspected to determine whether it contained a similar vulnerability, capturing semantically similar but structurally divergent cases.

The metrics used in previous RQs (R1, R10, and MRR) are not suitable here, as they assume a single ground truth per query. In contrast, this RQ addresses scenarios with multiple ground truths, necessitating a more comprehensive evaluation. Therefore, we adopt precision, recall, and F1 score, which are more appropriate for assessing performance in multi-ground truth settings.




\textbf{Results of homologous function detection.}
\tabl~\ref{tab:firmware} presents the homologous function detection results on the \vdf dataset. We report the performance of each tool in the top-10, top-25, and top-50 retrieved results. In addition, we apply a threshold-based approach to distinguish positive and negative samples, and report the threshold at which each tool achieves its highest F1 score in the last column. 
The results show that the proposed combination method consistently outperforms all individual tools across all evaluation settings, achieving an F1 score of 58.5\% in the top-25 results, representing a 13.4\% improvement over the best-performing individual tool, HermesSim (51.6\%).

\begin{tcolorbox}[colback=black!6!white,enhanced,frame hidden, boxsep=0pt,left=5pt,right=5pt,top=2pt,bottom=2pt]
\observation
The tool combination strategies demonstrate strong applicability and effectiveness (13.4\% improvement) in large-scale, real-world vulnerability detection scenarios.
\end{tcolorbox}

The F1 score at the top-25 cutoff consistently outperforms those at top-10, top-50, and the best-threshold settings. This can be attributed to the distribution of homologous functions per query, which has a median value of 23 and an average value of 27 that are both close to 25. Furthermore, the relatively poor performance under the best-threshold setting suggests that threshold-based classification is impractical in this context, as identifying the optimal threshold for each tool requires additional effort and tuning, limiting its applicability in real-world scenarios.



\textbf{Results of vulnerable homologous function detection.} 
Manual analysis of 500 confirmed homologous functions revealed that only 56\% (280) were actually vulnerable. This indicates that while BFSD tools effectively identify homologous functions, they struggle to distinguish between vulnerable and non-vulnerable ones. This limitation arises because BFSD tools emphasize overall functional similarity, whereas vulnerability detection often hinges on subtle differences in specific code fragments.
Additionally, manual verification is labor-intensive, requiring the identification of vulnerable locations from patches and matching them to corresponding code segments in target functions—an effort that varies with each vulnerability. 
This highlights the need for fine-grained techniques capable of matching code fragments within functions to approximate potential vulnerability locations. Such methods could substantially ease the verification burden and represent a promising avenue for future research.

\begin{tcolorbox}[colback=black!6!white,enhanced,frame hidden, boxsep=0pt,left=5pt,right=5pt,top=2pt,bottom=2pt]
\observation
BFSD tools cannot differentiate between vulnerable and non-vulnerable homologous functions. Effective vulnerability detection thus requires integration with fine-grained localization and validation techniques.
\end{tcolorbox}

\textbf{Results of vulnerable non-homologous function detection.}
We further analyzed 500 top-ranked non-homologous functions retrieved by the combination of HermesSim and \tool. Surprisingly, only one function exhibited the similar vulnerability to the queried ones-specifically, an incorrect return value check for a particular API. While some non-homologous functions shared functional similarities (e.g., image processing), this highlights a key limitation of BFSD tools: they can retrieve functionally similar but rarely vulnerability-equivalent functions. Addressing this challenge will likely require fundamentally new methodologies tailored to vulnerability semantics rather than general functional similarity.

\begin{tcolorbox}[colback=blue!5!white,enhanced,frame hidden, boxsep=0pt,left=5pt,right=5pt,top=2pt,bottom=2pt]
\insight
BFSD tools operate at the function level and often overlook fine-grained vulnerability semantics, limiting their effectiveness in detecting similar vulnerabilities that reside in small code regions.
\end{tcolorbox}

\section{Discussion}
\subsection{\add{Practical Implications for the Community}}

Our findings provide actionable strategies, practical guidance, and resources across multiple stakeholder groups.

\noindent\textbf{For practitioners and security analysts}, we recommend selecting and combining complementary BFSD tools based on different representations with comparable performance levels. HermesSim and \tool represent favorable choices given their consistently strong performance (\S~\ref{RQ1}). When constraints exist, such as lightweight model requirements or evaluating new tools, practitioners should identify target scenario characteristics (e.g., cross-architecture, cross-bitness), construct test datasets with \dataset under similar conditions, and apply the selection criteria established in this work.

\noindent\textbf{For researchers and tool developers}, our study identifies critical directions (\S ~\ref{sec:future_direction}). The insights on representation-level abstractions and tool complementarity also provide concrete guidance for designing next-generation BFSD techniques.

\noindent\textbf{For educators and the community}, we provide open-source datasets (\dataset and \vdf) and ready-to-use implementations, lowering barriers to reproducible research and accelerating progress in the BFSD field.

\subsection{Future Directions for BFSD}
\label{sec:future_direction}

\noindent\textbf{Addressing function inlining.}
As shown in \S~\ref{RQ1}, inconsistent inlining between homologous functions significantly disrupts semantics and leads to detection failures. Although prior work~\cite{Jia2024CrossInliningBF,ASM2Vec,CodeExtract} attempts to mitigate this, complex inlining remains challenging. Future work could explore partial matching or inclusion-based methods to better handle inlining-induced variability.

\noindent\textbf{Unifying multiple representations.}
Our findings in \S~\ref{RQ2} and \S~\ref{RQ3} show that tools based on distinct representations exhibit complementary strengths. Combining such tools yields notable performance gains, highlighting the promise of a unified BFSD framework that integrates multiple representations for enhanced robustness and accuracy.

\noindent\textbf{Scaling to large search spaces.}
Tool performance degrades with larger candidate pools (\S~\ref{RQ1}, \S~\ref{RQ3}), limiting scalability. Effective filtering is essential. Techniques like software composition analysis (SCA)\cite{Feng2019B2SFinderDO,Zhao2023UVSCANDT} can help narrow the search space. Application-specific filtering strategies, possibly integrating recent advances\cite{CEBin,AsteriaPro}, offer a promising path forward.

\noindent\textbf{Reducing manual verification effort.}
Current BFSD tools typically return similarity scores, requiring costly manual inspection to check if they are the searching targets. In \S~\ref{RQ3}, ground-truth labeling demanded substantial human effort. Two directions may help with this: (1) fine-grained matching to localize relevant code fragments, and (2) leveraging LLMs for automated result validation. 
Our preliminary experiments suggest that LLMs can effectively validate both homologous and vulnerable function matches. We provide the results of the preliminary experiment on our website.

\noindent\textbf{Towards finer-grained search.}
BFSD tools focus on global similarity and struggle to pinpoint specific vulnerabilities. This limits their effectiveness in tasks like vulnerability detection, where identifying precise vulnerable code regions is critical. Future research should explore searching at a finer granularity—e.g., matching vulnerable code fragments or semantic patterns within functions—beyond coarse-grained function similarity.

\add{
\noindent\textbf{Leveraging higher-level representations.}
Our evaluation reveals that top-performing tools, \tool\ and HermesSim, share a key design principle: lifting binary functions to unified, higher-level representations. \tool\ employs decompiled code, while HermesSim utilizes intermediate representation and constructs semantic-oriented graphs through static analysis. These abstractions effectively eliminate architectural differences and reduce inconsistencies, unlike other tools operating directly on assembly code or dependency graphs. This suggests that future BFSD research should prioritize architecture-agnostic, higher-level representations to achieve robust cross-platform and cross-optimization detection.
}

\subsection{Threats to Validity}

\noindent\textbf{External validity.}
Our findings are derived from experiments on two datasets, \dataset and \vdf. However, their generalizability to other settings, such as obfuscated code~\cite{Luo2014SemanticsbasedOB, ASM2Vec, zhang2023khaos} or proprietary binaries, remains unverified. Broader validation on larger and more diverse datasets is needed.
While our evaluation covers a subset of available tools, the tool combination strategies, and the limitations of BFSD tools are tool-agnostic and applicable across BFSD frameworks. \add{Moreover, our tools are trained on the non-inlined dataset. While training on the inlined dataset may yield performance improvements, such improvements may not generalize to all inlining patterns, as these patterns are content-dependent and vary across different functions.}



\noindent\textbf{Internal validity.}
Our experiments face two primary threats to internal validity. First, in \S~\ref{RQ1} and \S~\ref{RQ2}, query functions were selected randomly, which may introduce variability. To address this, we used 1,000 query functions and reported averaged results. Second, in \S~\ref{RQ3}, results required manual verification, which may involve subjectivity or errors. To reduce this risk, we adopted a two-person verification protocol: one performed the analysis, and a second independently reviewed and confirmed the results.

\section{Related Works}

\subsection{Evaluation Studies of BFSD Tools}
Recent studies have evaluated BFSD tools in different settings.
Marcelli et al.\cite{CiscoStudy} compared fuzzy hashing and machine learning approaches, revealing that machine learning methods performs better. Fu et al.\cite{AIPoweredStudy} assessed AI-based BFSD methods and their downstream applications, providing a structural analysis of common neural networks-based approaches.
These prior works exhibit three key limitations:
\textbf{Limited evaluation scope.} Existing studies often overlook critical real-world factors, such as function inlining and varying pool sizes, thereby limiting the depth and generalizability of their findings.
\textbf{Lack of analysis on practical usage.} Prior evaluations focus primarily on assessing the effectiveness of individual tools, missing the opportunity to explore practical usage strategies, such as combining tools or adapting them to specific contexts, that could improve real-world applicability.
\textbf{Non-representative and biased datasets.} Many evaluations rely on small-scale datasets, typically involving fewer than 10 vulnerabilities or firmware images, with limited architectural and project diversity. Furthermore, some datasets suffer from issues such as mislabeling (e.g., misclassified inlined functions) and heavy project bias (e.g., 57.6\% of functions sourced from Z3), undermining the reliability and generalizability of the results.

In contrast, our evaluation offers more reliable and representative insights, enabled by higher-quality and more diverse datasets. Particularly, we propose an actionable strategy for tool combination, which shows practical effectiveness.

\subsection{Literature Reviews of BFSD Approaches}
Several surveys have reviewed BFSD methods over time~\cite{haq2021survey, alrabaee2022survey, ruan2024survey}. Haq et al.\cite{haq2021survey} provided a broad overview of BFSD techniques, including their characteristics, implementations, and applications. Alrabaee et al.\cite{alrabaee2022survey} focused on binary code fingerprinting, categorizing methods by similarity levels, features, and detection strategies. Ruan et al.~\cite{ruan2024survey} offered a multidimensional comparison, analyzing the strengths and limitations in addressing the diverse characteristics of binary code. However, these surveys primarily offer qualitative analyses and lack quantitative evaluations of tool performance.

\section{Conclusion}
In this paper, we present the first large-scale empirical study of AI-based BFSD tools on two high-quality and diverse datasets. Based on three research questions, we propose an actionable tool combination strategy, that demonstrates strong effectiveness in large-scale, real-world vulnerability detection tasks, with an improvement of 13.4\%. Furthermore, we identify key limitations of current BFSD tools and offer insights for future research, including the promising potential of LLMs in enhancing BFSD capabilities.

\section*{Acknowledgment}
The authors would like to thank the anonymous reviewers for their helpful feedback on an earlier version of this paper. 
This work is partly supported by National Key R\&D Program of China under Grant \#2022YFB3103901, Chinese National Natural Science Foundation (Grants \#62202462, \#62302500, \#62032010). 
This research is supported by the National Research Foundation, Singapore, and DSO National Laboratories under the AI Singapore Programme (AISG Award No: AISG4-GC-2023-008-1B); by the National Research Foundation Singapore and the Cyber Security Agency under the National Cybersecurity R\&D Programme (NCRP25-P04-TAICeN); and by the Prime Minister’s Office, Singapore under the Campus for Research Excellence and Technological Enterprise (CREATE) Programme.
Any opinions, findings and conclusions, or recommendations expressed in these materials are those of the author(s) and do not reflect the views of the National Research Foundation, Singapore, Cyber Security Agency of Singapore, Singapore.


\bibliographystyle{IEEEtran}
\bibliography{ref}


\end{document}